\documentclass[12pt]{article}

\input tcilatex
\QQQ{Language}{
American English
}

\begin{document}

\author{I. Radinschi \and ``Gh. Asachi'' Technical University, Department of
Physics, \and B-dul D. Mangeron No.67, Iasi, 6600, Romania \and %
jessica@etc.tuiasi.ro}
\title{The Energy Distribution of the Bianchi Type I Universe}
\date{}
\maketitle

\begin{abstract}
We calculate the energy distribution of an anisotropic model of universe
based on the Bianchi type I metric in the Tolman's prescription. The energy
due to the matter plus gravitational field is equal to zero. This result
agrees with the results of Banerjee and Sen and Xulu. Also, our result
supports the viewpoint of Tryon and Rosen.


\end{abstract}

\section{Introduction}

The energy-momentum localization has been a problematic issue since the
outset of the theory of relativity. A large number of definitions of the
gravitational energy have been given since now. Some of them are coordinate
independent and other are coordinate-dependent. An adequate
coordinate-independent prescription for energy-momentum localization for all
the type of space-times has not given yet in General Relativity. We remark
that it is possible to evaluate the energy and momentum distribution by
using various energy-momentum complexes. There prevails suspicion that
different energy-momentum complexes could give different energy
distributions in a given space-time. Virbhadra and his collaborators [1]
have considered many space-times and have shown that several energy-momentum
complexes give the same and acceptable result for a given space-time. Also,
in his recent paper Virbhadra [2] emphasized that though the energy-momentum
complexes are non-tensors under general coordinate transformations, the
local conservation laws with them hold in all coordinate systems.

The subject of the energy of the Universe was re-opened by Cooperstock [3]
and Rosen [4]. Rosen [4] computed in the Einstein's prescription the energy
of a closed homogeneous isotropic universe described by a
Friedmann--Robertson--Walker (FRW) metric. The total energy is zero. Johri
[5], by using the Landau and Lifshitz energy-momentum complex, found that
the total energy of a FRW spatially closed universe is zero at all times
irrespective of equations of state of the cosmic fluid. Also, the total
energy enclosed within any finite volume of the spatially flat FRW universe
is zero at all times. It is known that the Bianchi type I solutions, under a
special case, reduce to the spatially flat FRW solutions. The total energy
density is found to be zero everywhere. Banerjee and Sen [6] calculated in
the Einstein's prescription the total energy density of the Bianchi type I
solutions. Also, Xulu [7], by using the Landau and Lifshitz, Papapetrou and
Weinberg prescriptions found that the total energy of the Universe in the
case of the Bianchi type I model is zero.

The purpose of this paper is to compute the energy of an anisotropic model
of universe based on the Bianchi type I metric by using the energy-momentum
complex of Tolman. We also make a discussion of the results. We use the
geometrized units ($G=1,c=1$) and follow the convention that the Latin
indices run from 0 to 3.

\section{The energy in the Tolman's prescription}

We consider the line element [8] which describes a special anisotropic model
of universe based on the Bianchi type I metric

\begin{equation}
ds^2=dt^2-A^2(t)dx^2-B^2(t)dy^2-C^2(t)dz^2,
\end{equation}
where

\begin{equation}
\begin{array}{l}
A(t)=(m_1s_1t)^{1/m_1}, \\[6pt] 
B(t)=(m_2s_2t)^{1/m_2}, \\[6pt] 
C(t)=(m_3s_3t)^{1/m_3}.
\end{array}
\end{equation}
In (2) $m_i$ and $s_i$ ($i=\overline{1,3}$) are positive constants and we
exclude the $m_i=0$ case.

The metric given by (1) reduces to the spatially flat
Friedmann--Robertson--Walker metric in a special case when we have $%
A(t)=B(t)=C(t)=R(t)$. We define $R(t)=(mst)^{1/m}$ and transforming the line
element (1) according to

\begin{equation}
\cases{x=r\sin\theta\cos\varphi,\cr y=r\sin\theta\sin\varphi,\cr
z=r\cos\theta.}
\end{equation}

We obtain the line element

\begin{equation}
ds^2=dt^2-R^2(t)[dr^2+r^2(d\theta ^2+\sin ^2\theta d\varphi ^2)],
\end{equation}
which describes the spatially flat Friedmann--Robertson--Walker space-time.

The only non-zero components of the energy-momentum tensor (due to the
matter) are

\begin{equation}
\begin{array}{l}
\displaystyle T_1^{\;1}=\left( {\frac{m_2-1}{m_2^2}}+{\frac{m_3-1}{m_3^2}}-{%
\frac 1{m_2m_3}}\right) t^{-2}, \\[12pt] 
\displaystyle T_2^{\;2}=\left( {\frac{m_1-1}{m_1^2}}+{\frac{m_3-1}{m_3^2}}-{%
\frac 1{m_1m_3}}\right) t^{-2}, \\[12pt] 
\displaystyle T_3^{\;3}=\left( {\frac{m_1-1}{m_1^2}}+{\frac{m_2-1}{m_2^2}}-{%
\frac 1{m_1m_2}}\right) t^{-2}, \\[12pt] 
\displaystyle T_0^{\;0}=\left( {\frac{m_1+m_2+m_3}{m_1m_2m_3}}\right) t^{-2}.
\end{array}
\end{equation}

From (5) it results that the energy density component of the energy-momentum
tensor is not zero for the Bianchi type I solutions.

The Tolman's energy-momentum complex [9] is given by

\begin{equation}
\Upsilon _i^{\;\;k}={\frac 1{8\pi }}U_i^{\;kl},_l,
\end{equation}
where $\Upsilon _0^{\;\;0}$ and $\Upsilon _\alpha ^{\;\;0}$ are the energy
and momentum components.

We have

\begin{equation}
U_i^{\;kl}=\sqrt{-g}(-g^{pk}V_{ip}^{\;~l}+\frac
12g_i^kg^{pm}V_{pm}^{\;~\;\;l}),
\end{equation}
with

\begin{equation}
V_{jk}^{\;\;i}=-\Gamma _{jk}^i+{\frac 12}g_j^i\Gamma _{mk}^m+{\frac 12}%
g_k^i\Gamma _{mj}^m.
\end{equation}

The energy-momentum complex $\Upsilon _i^{\;\;k}$ also satisfies the local
conservation laws

\begin{equation}
\frac{\partial \Upsilon _i^{\;\;k}}{\partial x^k}=0.
\end{equation}

The energy and momentum in the Tolman's prescription are given by

\begin{equation}
P_i=\int \hskip-7pt\int \hskip-7pt\int \Upsilon _i^{\;\;0}dx^1dx^2dx^3.
\end{equation}

Using the Gauss's theorem we obtain

\begin{equation}
P_i={\frac 1{8\pi }}\int \hskip-7pt\int U_i^{\;0\alpha }n_\alpha dS,
\end{equation}
where $n_\alpha =(x/r,y/r,z/r)$ are the components of a normal vector over
an infinitesimal surface element $dS=r^2\sin \theta d\theta d\varphi $.

By using (6) and (7) we have that all the $U_0^{\;0i}$ components vanish and
we obtain

\begin{equation}
\Upsilon _0^{\;0}=0.
\end{equation}

The total energy density (due to the matter plus field) vanishes everywhere.

\section{Discussion}

The main purpose of the present paper is to show that it is possible to
''solve'' the problem of the localization of energy in relativity by using
the energy-momentum complexes.

The subject of the localization of energy continues to be an open one. Bondi
[10] sustained that a nonlocalizable form of energy is not admissible in
relativity. A favorable argument for using the energy-momentum complexes to
calculate the energy distribution of different models of universe is that
they can give the same result for a given space-time. Chang, Nester and Chen
[11] showed that the energy-momentum complexes are actually quasilocal and
legitimate expressions for the energy-momentum.

As we noted, Rosen [4] found that the total energy of a closed homogeneous
isotropic universe described by a FRW metric is zero. Johri [5] showed that
the total energy of a FRW spatially closed universe is zero at all times
irrespective of equations of state of the cosmic fluid. Also, the total
energy enclosed within any finite volume of the spatially flat FRW universe
is zero at all times. Banerjee and Sen [6] and Xulu [7] obtained that the
total energy density of a model of universe based on the Bianchi type I
solutions is zero.

We used a special case of the Bianchi type I metric and obtained a result
which agrees with the results of Banerjee and Sen [6] and Xulu [7]. The
total energy density vanishes everywhere because the energy contributions
from the matter and field inside an arbitrary two-surface, in the case of
the anisotropic model based on the Bianchi type I metric, cancel each other.
Our result supports the viewpoint of Tryon [12] which assumed that the
Universe appeared from nowhere about $10^{10}$ years ago and the
conventional laws of physics need not have been violated at the time of the
creation of the Universe. According with his Big Bang model, the Universe is
homogeneous, isotropic and closed and consists of matter and anti-matter
equally. Also, the net energy of the Universe may be equal to zero. Also,
our result supports the calculations of Rosen [4] which agrees with the
studies of Tryon. We use the Bianchi I space-time because, as we showed
previously, in this case, by using an adequate transformation of coordinates
we can reach the line element which describes the spatially flat
Friedmann--Robertson--Walker space-time. For an universe described by a
metric that can be reduce to the spatially flat FRW metric the total energy
is zero [5]-[7].

We completed the investigation of Banerjee and Sen [6] and Xulu [7] with one
more energy-momentum complex. The result in this paper sustains the
importance of the energy-momentum complexes.

\end{document}